\documentclass[superscriptaddress,showpacs,amssymb,amsmath,
amsfonts,aps,
prl,twocolumn
]{revtex4}
\usepackage{graphicx}
\usepackage{dcolumn}
\usepackage{epsfig}
\usepackage{subfigure}
\begin{document}

\title{Electromagnetic Decay of the $\Sigma^{0}(1385)$ to $\Lambda\gamma$}

\newcommand{\bra}[1]{\left\langle #1 \right|}
\newcommand{\ket}[1]{\left| #1 \right\rangle}
\newcommand{\bracket}[2]{\left\langle #1 | #2 \right\rangle}
\newcommand{\matrixelement}[3]{\bra{#1} \hat{#2} \ket{#3}}
\newcommand\polvec{\vec\epsilon_\lambda^{\,*}(\vec k)}
\newcommand\etal{{\em et al.}}

\pacs{13.40.Em,14.20.Jn,13.30.Ce,13.40.Hq}

\newcommand*{\ANL}{Argonne National Laboratory, Argonne, Illinois 60441}
\newcommand*{\ANLindex}{1}
\affiliation{\ANL}
\newcommand*{\ASU}{Arizona State University, Tempe, Arizona 85287-1504}
\newcommand*{\ASUindex}{2}
\affiliation{\ASU}
\newcommand*{\CSUDH}{California State University, Dominguez Hills, Carson, CA 90747}
\newcommand*{\CSUDHindex}{3}
\affiliation{\CSUDH}
\newcommand*{\CANISIUS}{Canisius College, Buffalo, NY}
\newcommand*{\CANISIUSindex}{4}
\affiliation{\CANISIUS}
\newcommand*{\CMU}{Carnegie Mellon University, Pittsburgh, Pennsylvania 15213}
\newcommand*{\CMUindex}{5}
\affiliation{\CMU}
\newcommand*{\CUA}{Catholic University of America, Washington, D.C. 20064}
\newcommand*{\CUAindex}{6}
\affiliation{\CUA}
\newcommand*{\SACLAY}{CEA, Centre de Saclay, Irfu/Service de Physique Nucl\'eaire, 91191 Gif-sur-Yvette, France}
\newcommand*{\SACLAYindex}{7}
\affiliation{\SACLAY}
\newcommand*{\CNU}{Christopher Newport University, Newport News, Virginia 23606}
\newcommand*{\CNUindex}{8}
\affiliation{\CNU}
\newcommand*{\UCONN}{University of Connecticut, Storrs, Connecticut 06269}
\newcommand*{\UCONNindex}{9}
\affiliation{\UCONN}
\newcommand*{\EDINBURGH}{Edinburgh University, Edinburgh EH9 3JZ, United Kingdom}
\newcommand*{\EDINBURGHindex}{10}
\affiliation{\EDINBURGH}
\newcommand*{\FU}{Fairfield University, Fairfield CT 06824}
\newcommand*{\FUindex}{11}
\affiliation{\FU}
\newcommand*{\FIU}{Florida International University, Miami, Florida 33199}
\newcommand*{\FIUindex}{12}
\affiliation{\FIU}
\newcommand*{\FSU}{Florida State University, Tallahassee, Florida 32306}
\newcommand*{\FSUindex}{13}
\affiliation{\FSU}
\newcommand*{\Genova}{Universit$\grave{a}$ di Genova, 16146 Genova, Italy}
\newcommand*{\Genovaindex}{14}
\affiliation{\Genova}
\newcommand*{\GWUI}{The George Washington University, Washington, DC 20052}
\newcommand*{\GWUIindex}{15}
\affiliation{\GWUI}
\newcommand*{\ISU}{Idaho State University, Pocatello, Idaho 83209}
\newcommand*{\ISUindex}{16}
\affiliation{\ISU}
\newcommand*{\INFNFE}{INFN, Sezione di Ferrara, 44100 Ferrara, Italy}
\newcommand*{\INFNFEindex}{17}
\affiliation{\INFNFE}
\newcommand*{\INFNFR}{INFN, Laboratori Nazionali di Frascati, 00044 Frascati, Italy}
\newcommand*{\INFNFRindex}{18}
\affiliation{\INFNFR}
\newcommand*{\INFNGE}{INFN, Sezione di Genova, 16146 Genova, Italy}
\newcommand*{\INFNGEindex}{19}
\affiliation{\INFNGE}
\newcommand*{\INFNRO}{INFN, Sezione di Roma Tor Vergata, 00133 Rome, Italy}
\newcommand*{\INFNROindex}{20}
\affiliation{\INFNRO}
\newcommand*{\ORSAY}{Institut de Physique Nucl\'eaire ORSAY, Orsay, France}
\newcommand*{\ORSAYindex}{21}
\affiliation{\ORSAY}
\newcommand*{\ITEP}{Institute of Theoretical and Experimental Physics, Moscow, 117259, Russia}
\newcommand*{\ITEPindex}{22}
\affiliation{\ITEP}
\newcommand*{\JMU}{James Madison University, Harrisonburg, Virginia 22807}
\newcommand*{\JMUindex}{23}
\affiliation{\JMU}
\newcommand*{\KNU}{Kyungpook National University, Daegu 702-701, Republic of Korea}
\newcommand*{\KNUindex}{24}
\affiliation{\KNU}
\newcommand*{\LPSC}{LPSC, Universite Joseph Fourier, CNRS/IN2P3, INPG, Grenoble, France}
\newcommand*{\LPSCindex}{25}
\affiliation{\LPSC}
\newcommand*{\UNH}{University of New Hampshire, Durham, New Hampshire 03824-3568}
\newcommand*{\UNHindex}{26}
\affiliation{\UNH}
\newcommand*{\NSU}{Norfolk State University, Norfolk, Virginia 23504}
\newcommand*{\NSUindex}{27}
\affiliation{\NSU}
\newcommand*{\OHIOU}{Ohio University, Athens, Ohio  45701}
\newcommand*{\OHIOUindex}{28}
\affiliation{\OHIOU}
\newcommand*{\ODU}{Old Dominion University, Norfolk, Virginia 23529}
\newcommand*{\ODUindex}{29}
\affiliation{\ODU}
\newcommand*{\RPI}{Rensselaer Polytechnic Institute, Troy, New York 12180-3590}
\newcommand*{\RPIindex}{30}
\affiliation{\RPI}
\newcommand*{\ROMAII}{Universita' di Roma Tor Vergata, 00133 Rome Italy}
\newcommand*{\ROMAIIindex}{31}
\affiliation{\ROMAII}
\newcommand*{\MSU}{Skobeltsyn Nuclear Physics Institute, Skobeltsyn Nuclear Physics Institute, 119899 Moscow, Russia}
\newcommand*{\MSUindex}{32}
\affiliation{\MSU}
\newcommand*{\SCAROLINA}{University of South Carolina, Columbia, South Carolina 29208}
\newcommand*{\SCAROLINAindex}{33}
\affiliation{\SCAROLINA}
\newcommand*{\JLAB}{Thomas Jefferson National Accelerator Facility, Newport News, Virginia 23606}
\newcommand*{\JLABindex}{34}
\affiliation{\JLAB}
\newcommand*{\UNIONC}{Union College, Schenectady, NY 12308}
\newcommand*{\UNIONCindex}{35}
\affiliation{\UNIONC}
\newcommand*{\UTFSM}{Universidad T\'{e}cnica Federico Santa Mar\'{i}a, Casilla 110-V Valpara\'{i}so, Chile}
\newcommand*{\UTFSMindex}{36}
\affiliation{\UTFSM}
\newcommand*{\GLASGOW}{University of Glasgow, Glasgow G12 8QQ, United Kingdom}
\newcommand*{\GLASGOWindex}{37}
\affiliation{\GLASGOW}
\newcommand*{\VT}{Virginia Polytechnic Institute and State University, Blacksburg, Virginia   24061-0435}
\newcommand*{\VTindex}{38}
\affiliation{\VT}
\newcommand*{\VIRGINIA}{University of Virginia, Charlottesville, Virginia 22901}
\newcommand*{\VIRGINIAindex}{39}
\affiliation{\VIRGINIA}
\newcommand*{\WM}{College of William and Mary, Williamsburg, Virginia 23187-8795}
\newcommand*{\WMindex}{40}
\affiliation{\WM}
\newcommand*{\YEREVAN}{Yerevan Physics Institute, 375036 Yerevan, Armenia}
\newcommand*{\YEREVANindex}{41}
\affiliation{\YEREVAN}

\newcommand*{\NOWMSU}{Skobeltsyn Nuclear Physics Institute, Skobeltsyn Nuclear Physics Institute, 119899 Moscow, Russia}
\newcommand*{\NOWINFNGE}{INFN, Sezione di Genova, 16146 Genova, Italy}

\author {D.~Keller} 
\affiliation{\OHIOU}
\author {K.~Hicks} 
\affiliation{\OHIOU}
\author {K.P. ~Adhikari} 
\affiliation{\ODU}
\author {K.P. ~Adhikari} 
\affiliation{\ODU}
\author {D.~Adikaram} 
\affiliation{\ODU}
\author {M.~Aghasyan} 
\affiliation{\INFNFR}
\author {M.~Amarian} 
\affiliation{\ODU}
\author {H.~Baghdasaryan} 
\affiliation{\VIRGINIA}
\affiliation{\ODU}
\author {J.~Ball} 
\affiliation{\SACLAY}
\author {M.~Battaglieri} 
\affiliation{\INFNGE}
% ...

\collaboration{The CLAS Collaboration}
\noaffiliation
\date{\today}

\begin{abstract}
The electromagnetic decay $\Sigma^0(1385) \to \Lambda \gamma$
was studied using the CLAS detector at the 
Thomas Jefferson National Accelerator Facility. 
A real photon beam with a maximum energy of 3.8 GeV was 
incident on a proton target, producing an exclusive final 
state of $K^+\Sigma^{*0}$.
We report the decay widths ratio $\Sigma^0(1385) \to \Lambda\gamma$/ 
$\Sigma^0(1385) \to \Lambda\pi^0$ = $1.42 \pm 0.12(\text{stat})_{-0.07}^{+0.11}(\text{sys})$\%. 
This ratio is larger than most theoretical predictions by factors ranging from 1.5-3, but is
consistent with the only other experimental measurement.  From the reported ratio
we calculate the partial width and electromagnetic transition magnetic moment for 
$\Sigma^0(1385) \to \Lambda\gamma$.
\end{abstract}

\maketitle

\section{Introduction}

One well-known success of the constituent quark model (CQM) 
is its prediction of the low-mass 
baryon magnetic moments, using just the SU(6) wavefunctions \cite{Beg,Rubin}.
Calculations of the magnetic moments \cite{IsgKar}, assuming that 
quarks behave as point-like Dirac dipoles, are typically within 
$\sim$10\% of the current measured values \cite{PDG}.
However, today we know that the spin of the proton is much 
more complex than the CQM representation, with only about one-third of the proton's spin 
coming from the quarks and the rest of the 
spin resulting from a combination of the gluon spins and the orbital angular momentum 
of the quarks \cite{Alexakhin,Airapetian}. Clearly, the CQM is an over-simplification 
of the spin dynamics inside baryons, yet somehow the CQM captures 
the degrees of freedom that are relevant to the measured magnetic moments.
Further measurements of baryon magnetic 
moments, utilizing the electromagnetic decays of excited baryons, will 
continue to test our understanding of baryon wavefunctions.

Experimentally, it is difficult to measure the electromagnetic
(EM) transitions of decuplet-to-octet baryons because of the competition 
between EM and strong decays. For 
example, the branching ratio for EM decay of the $\Delta$ resonance 
has been measured to be about 0.55\% \cite{PDG} and branching ratios 
for other decuplet baryons are predicted to be of the same order 
of magnitude.
The EM transition form factors for the $\Delta$ may
be directly measured via pion photoproduction \cite{Dalitz,Sato}. 

It has been shown \cite{Lee} that pion cloud effects 
contribute significantly ($\sim$40\%) to the $\gamma p \to \Delta^+$  
magnetic dipole transition form factor, $G_M(Q^2)$, at low $Q^2$ 
(below $\sim 0.1$ GeV$^2$).  In the naive non-relativistic quark 
model \cite{CQM}, the value of $G_M(0)$ is directly proportional to 
the proton magnetic moment, and measurements of $G_M$ 
near $Q^2=0$ can only be explained (within this quark model) if 
the experimental magnetic moment is lowered by about 30\%. 
This again suggests that the CQM is an over-simplification of 
reality.

To extend these measurements to the other decuplet baryons, 
which have non-zero strangeness, hyperons must be 
produced through strangeness-conserving reactions.  
Then their EM decay, which has a small branching ratio, must 
be measured directly. Although these measurements are difficult,
it is important to measure the EM decays of strange baryons 
to extract information on their wavefunctions, which in turn,
constrains theoretical models of baryon structure.
The measurements of EM transition form factors for 
decuplet baryons with strangeness may also be sensitive to meson 
cloud effects.  A comparison of the EM decay measurements to 
predictions of quark models for decay of decuplet hyperons, $\Sigma^*$, to octet hyperons, $Y$,
can provide a measure of the importance of meson cloud 
diagrams in the $\Sigma^* \to Y\gamma$ transition.

Here, we present measurements of the EM decay 
$\Sigma^{*0} \to \Lambda \gamma$ normalized to the strong decay 
$\Sigma^{*0} \to \Lambda \pi^0$.  
The present results can be compared to previous
measurements of the $\Sigma^{*0}$ EM decay \cite{Taylor} (also from CLAS data) that had a larger uncertainty ($\sim$25\% statistical and $\sim$15\% 
systematic uncertainty).  The smaller uncertainties here are 
due to a larger data set (more than 10 times bigger) and, subsequently, 
better control over systematic uncertainties.  The reduced 
uncertainty is important because, as mentioned above, meson cloud 
effects are predicted to be on the order of $\sim$30-40\%. 
In order to quantify the 
effect of meson clouds for baryons with non-zero strangeness, 
it is desirable to keep measurement uncertainties below $\sim$10\%.

\begin{table*}[htb]
\caption{Theoretical predictions for the models referenced in the text and the experimental values for the electromagnetic decay widths (in keV).}
\begin{center}
\begin{tabular}{lccc}
Model 
&~~~~~$\Delta(1232) \to N\gamma$~~~~~
&~~~~~$\Sigma(1193) \to \Lambda \gamma$~~~~~
&~~~~~$\Sigma(1385) \to \Lambda \gamma$~~~~~ \\
\hline\hline
NRQM \cite{Koniuk,kaxiras,DHK}	& 360	& 8.6	& 273	\\
RCQM \cite{warns}   		&	& 4.1	& 267	\\
$\chi$CQM \cite{wagner} 		& 350 	& 	& 265   \\
MIT Bag \cite{kaxiras}  		& 	& 4.6	& 152	\\
Soliton \cite{Schat} 		& 	&	& 243	\\
Skyrme \cite{Abada,Haberichter}	& 309-326 & & 157-209 	\\
Algebraic model \cite{Bijker} 	& 341.5 & 8.6	& 221.3	\\
HB$\chi$PT \cite{butler}$^\dag$ 	& (670-790) &  & 290-470 \\
				&	&	&	\\
\hline
Experiment \cite{PDG} 		& 660$\pm$47 & 8.9$\pm$0.9 & 470$\pm$160 \\
\hline\hline
$^\dag$ Normalized to experiment for the $\Delta \to N\gamma$ range shown.
\end{tabular}   
\end{center}
\label{tab:widths}
\end{table*}

There are many theoretical calculations of the EM decays of 
decuplet hyperons such as: 
the non-relativistic quark model (NRQM) \cite{DHK,Koniuk}, 
a relativized constituent quark model (RCQM) \cite{warns}, 
a chiral constituent quark model ($\chi$CQM) \cite{wagner},
the MIT bag model \cite{kaxiras}, 
the bound-state soliton model \cite{Schat}, 
a three-flavor generalization of the Skyrme model that uses the 
collective approach \cite{Abada,Haberichter},
an algebraic model of hadron structure \cite{Bijker}, and 
heavy baryon chiral perturbation theory (HB$\chi$PT) \cite{butler}, 
among others.  Table \ref{tab:widths} summarizes the theoretical 
predictions and experimental branching ratios for the EM 
transitions of interest.

A comprehensive study of electromagnetic strangeness production 
has been undertaken using the CLAS detector at the Thomas Jefferson 
National Accelerator Facility.  Many data on ground-state hyperon 
photoproduction have already been published
\cite{mcnabb,bradford,mccracken} 
using data from the CLAS run group $g1$ and $g11$ data sets. 
The $g1$ experiment had an open trigger \cite{mcnabb} and a
lower data acquisition speed, whereas the $g11$ experiment required 
that at least two particles be detected \cite{mccracken}, and used a higher 
beam current, resulting in much higher data acquisition speed.
The result is that the $g11$ data set had over 20 times more 
reconstructed events than the $g1$ data set.  The present results use the 
$g11$ data set, whereas Taylor \etal~\cite{Taylor} used the $g1$ data set.
Published CLAS results \cite{mccracken} from $g11$ demonstrate the
accurate calibration of this data set, and that the cross section of
$\gamma p \to K^+ \Lambda$ matches previous CLAS data.

The EM decay of the $\Sigma^{*0}$ is only about 1\% of the 
total decay width. To isolate this signal from the dominant 
strong decay $\Sigma^{*0} \to \Lambda \pi^0$, the missing mass 
of the detected particles, $\gamma p \to K^+ \Lambda (X)$, is calculated.
Because of its proximity to the $\pi^{0}$ peak in the mass spectrum 
from strong decay, the EM decay signal is difficult to separate 
using simple peak-fitting methods.  
The strategy here is to understand and eliminate 
as much background as possible using standard kinematic cuts, 
and then use a kinematic fitting procedure for each channel.  
As described below, by varying the cut points on the confidence 
levels of each kinematic fit, the systematic uncertainty 
associated with the extracted ratio for EM decay can 
be quantitatively determined.  The increased statistics for the 
$g11$ data helps greatly to study and to determine the systematic uncertainty
associated with the measurement.

\section{The Experiment}

For the present measurements, a bremsstrahlung photon beam was produced from 
a 4.019 GeV electron beam, resulting in a photon energy range of 1.6-3.8 GeV.  
The photon energy was deduced from a magnetic spectrometer \cite{tagnim} 
that ``tagged" the electron with an energy resolution of $\sim 0.1\%$.  
A liquid-hydrogen target was used that was 40 cm long and placed such that 
the center of the target was 10 cm upstream from the center of CLAS.
As mentioned above, a trigger requiring two charged particles in coincidence with the tagged 
electron was used.  
The data acquisition recorded approximately 20 billion events.
Details of the experimental setup are given elsewhere 
\cite{mccracken,mecking}.

\subsection{Event Selection}

We selected events for the reaction $\gamma p \to K^{+}\Sigma^{*0}$, 
where the $\Sigma^{*0}$ decays with 87.0$\pm$1.5$\%$ probability to 
$\Lambda\pi^{0}$ and 1.3$\pm$0.4\% probability to $\Lambda\gamma$ \cite{PDG}. 
The $\Lambda$ then decays weakly with 63.9$\pm$0.5$\%$ probability to $p\pi^-$ \cite{PDG}, 
leading to the final states $\gamma p \to K^{+} p \pi^{-} \pi^{0}$ and 
$\gamma p \to K^{+} p \pi^{-} \gamma$, respectively.  
The charged particles are tracked by the CLAS drift chambers through
the magnetic field of the spectrometer, giving their 
momentum, and are detected by the time-of-flight scintillators, giving their velocity.
The drift chamber tracking covariance matrix is obtained for each track.  This
contains the uncertainty in each measured variable used in track reconstruction along
with the appropriate correlations. 
The $\pi^{0}$ and $\gamma$ must be deduced indirectly using conservation of 
energy and momentum via the missing mass technique.  

In the present analysis, two positively charged particles and one negatively charge particle
are selected.  The mass of the detected particles was 
calculated from the measured velocity and momentum.  
The mass is given by 
\begin{equation}
m_{\text{cal}} = \sqrt{{p^{2}(1 - \beta^{2}) \over \beta^{2}}},
\end{equation}
where $\beta = L/t_{\text{meas}}$ for path length $L$ and measured 
time-of-flight $t_{\text{meas}}$, and the speed of light is set to 1.
The pions, kaons, and protons were identified using mass cuts of
$0.0\leq M_{\pi^{-}} \leq 0.3$ GeV, $0.3 < M_{K^+} < 0.8$ GeV, and 
$0.8\leq M_{p} \leq 1.2$ GeV, respectively.  
From this initial identification it is possible to incorporate 
additional timing information to improve event selection.  
The time-of-flight $t_{\text{meas}}$ is the time difference 
between the event vertex time and the time at which the particle 
strikes the time-of-flight scintillator wall at the outside 
of the CLAS detector. We define $\Delta t = t_{\text{meas}} - t_{\text{cal}}$, 
where $t_{\text{cal}}$ is the time-of-flight calculated for an assumed 
mass such that 
\begin{equation}
t_{\text{cal}} = L \sqrt{1 + { \left(m \over p \right)}^{2}},
\end{equation}
where $m$ is the assumed mass for the particle of interest and $p$ is the
momentum magnitude.  A cut on $\Delta t$ or $m_{\text{cal}}$ should be 
effectively equivalent.

Using $\Delta t$ for each particle it is possible to reject events that 
are not associated with the correct RF beam bunch, which are separated 
by 2 ns.  This is done by accepting only events with $|\Delta t| \leq$1 ns.  

A $\Delta \beta$ cut also helps to clean up the identification scheme.
$\Delta \beta$ is the difference between the above measured $\beta$
and the calculated $\beta_c$ defined by $\beta_c=p/\sqrt{p^2+m^2}$, where
$p$ is the particle momentum and $m$ is the known particle mass.
The good events were required to have $-0.02\le \Delta \beta \le 0.02$.

\begin{figure}
\epsfig{file=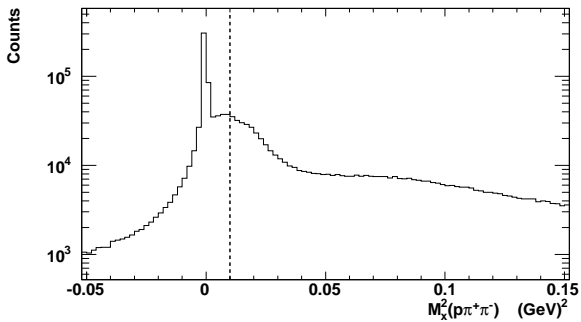,width=\columnwidth}
\caption{Missing mass squared $(M^2_{x})$ for the reaction 
$\gamma p \to p \pi^{+} \pi^{-} (X)$, where the $\pi^{+}$ was 
a potentially misidentified kaon.  Events above the dotted line at 0.01 GeV$^2$ 
were kept.}
\label{T1}
\end{figure}
 
The energy lost by charged particles passing through the CLAS detector was
accounted for by adjusting the measured particle's energy according to the
average $dE/dx$ losses in the target material, target wall, target scattering chamber, and
the start counter scintillators surrounding the target.
After correcting for energy loss, several kinematic cuts are applied 
as described below.
\begin{figure}
\epsfig{file=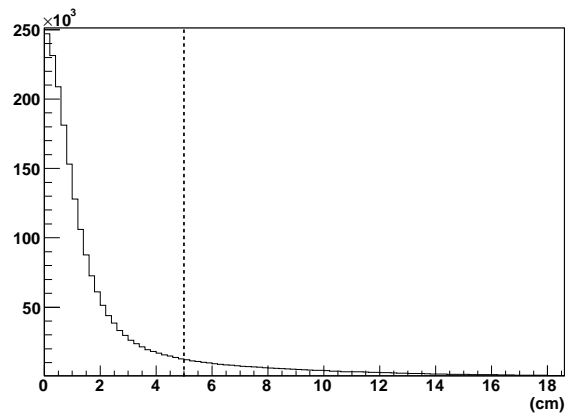,width=\columnwidth}
\caption{The distance of closest approach between the proton and $\pi^-$
shown in cm.  The cut used at 5 cm is indicated by the dotted line.}
\label{doca}
\end{figure}

\begin{figure}
\epsfig{file=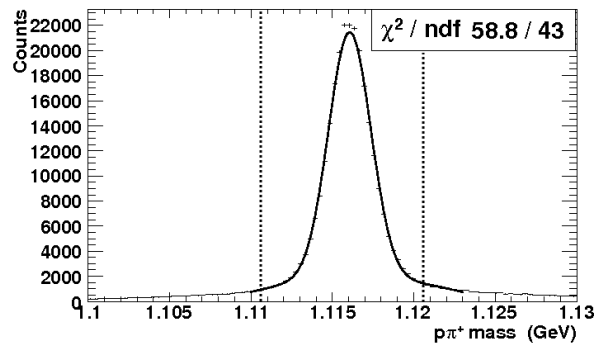,width=\columnwidth}
\caption{The invariant mass of the $p$-$\pi^{-}$ showing the $\Lambda$ peak
with a Gaussian fit giving a $\sigma=1.3$ MeV.  A cut of $\pm0.005$ GeV around
the PDG mass of the $\Lambda$ is used indicated by the dotted lines.}
\label{lam}
\end{figure}

Due to the finite resolution of the measured velocity and momentum, 
in addition to particle decay-in-flight, it is possible that some pions 
could be misidentified as kaons.  
To clean up the kaon signal for the analysis, it is common to 
recalculate the energy of the identified kaon using the mass of the pion.
Then the missing mass squared is studied for the reaction 
$\gamma p \rightarrow p \pi^+\pi^-(X)$, where the $\pi^+$ is actually 
identified by the above mass cuts as a $K^+$.
A spike at zero mass squared indicates that the reaction 
$\gamma p \rightarrow p \pi^+\pi^-$ is prominent.  
Most particle misidentified events can be removed by cutting slightly 
above zero, as shown in Fig. \ref{T1}.  
The events above 0.01 GeV$^{2}$ are kept as a cleaner sample of the $\gamma p \rightarrow p K^+\pi^-(X)$ events.  
Reactions involving decays such as $\rho \to \pi^{+} \pi^{-}$, where the $\pi^+$ 
is mistakenly identified as a $K^+$, are vastly reduced by this cut.  

\begin{figure}
\epsfig{file=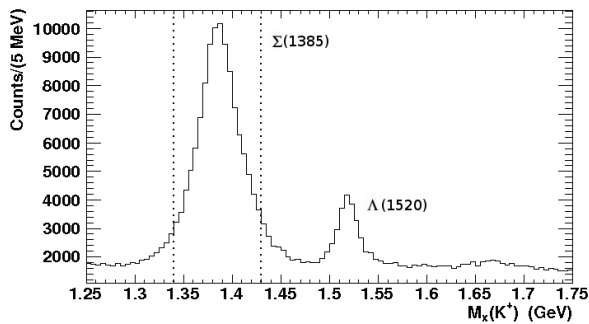,width=\columnwidth}
\caption{Missing mass for the reaction $\gamma p \to K^{+}(X)$,
for events passing the cut on the $\Lambda$ mass.  The dotted lines
show the 1.34 GeV to 1.43 GeV cut used to select the $\Sigma(1385)$.}
\label{offKp}
\end{figure}

The four-momentum of the detected $\Lambda$ was reconstructed from the
proton and $\pi^{-}$ four-momenta.  The distance of
closest approach (DOCA) from the proton and $\pi^{-}$ four-momenta is found
and restricted to be less than 5 cm (see Fig. \ref{doca}).
A Gaussian fit to the $p\pi^-$ invariant mass peak shown in
Fig. \ref{lam} resulted in 
a $\sigma = 1.3$ MeV, which is consistent with the instrumental resolution.  
After restricting the $\Lambda$ mass to be $1.1157\pm0.005$ GeV,
the remaining events were used to construct the missing mass off the $K^+$, giving the excited-state hyperon mass spectrum shown in Fig. \ref{offKp}.

\begin{figure}
\epsfig{file=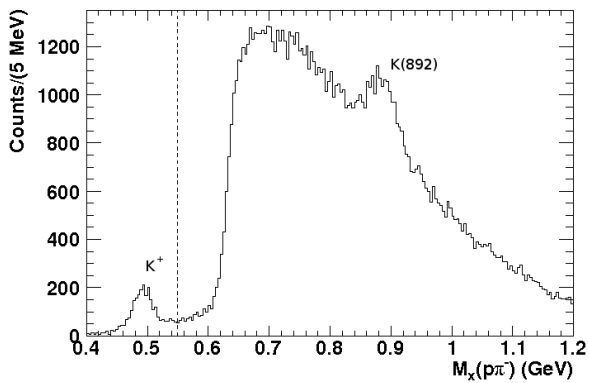,width=\columnwidth}
\caption{Missing mass for the reaction $\gamma p \to \Lambda(X)$ 
for events passing cuts on the $\Lambda$ and $\Sigma^*$ masses.
The dotted line at 0.55 GeV shows the cut used to remove the $\gamma p \to K^+\Lambda$ channel.
(A looser timing cut was used to illustrate that these accidentals are at the $K^+$
mass.)}
\label{offLam}
\end{figure}

After making a cut on the $\Sigma^*$ peak from 1.34-1.43 GeV, as shown in Fig. \ref{offKp}, 
one can study the missing mass off of the $\Lambda$, such that $\gamma p \to \Lambda (X)$, shown in 
Fig. \ref{offLam}.  Small peaks are seen at the mass of the kaon and 
the $K^{*}(892)$.
The kaon peak is from exclusive $\gamma p \to K^+ \Lambda$ production 
due to accidentals under the TOF peak, and can easily be cut out.
The dotted line shows the $M_x(\Lambda) > 0.55$ GeV event 
selection used to eliminate this background.

\begin{figure}
\epsfig{file=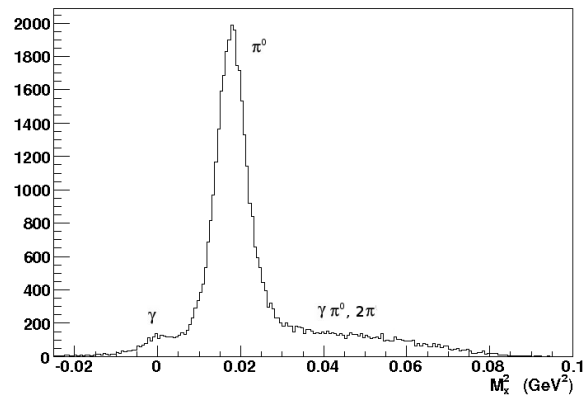,width=\columnwidth}
\caption{Missing mass squared for the reaction 
$\gamma p \to K^{+} p\pi^{-}(X)$ after all kinematic cuts. }
\label{MM1}
\end{figure}

After including all of the cuts listed above, the missing mass of the reaction 
$\gamma p \to K^{+} \Lambda (X)$ is shown in Fig. \ref{MM1}.
A very prominent peak is seen at the mass of the $\pi^{0}$ with 
a very small number of counts about zero missing mass due to the EM decay.  
The counts above the $\pi^{0}$ peak are mostly due to the 
$\gamma p \to K^{+} \Sigma^{0}(X)$ reaction from photoproduction 
of higher-mass hyperons.

\begin{figure}
\epsfig{file=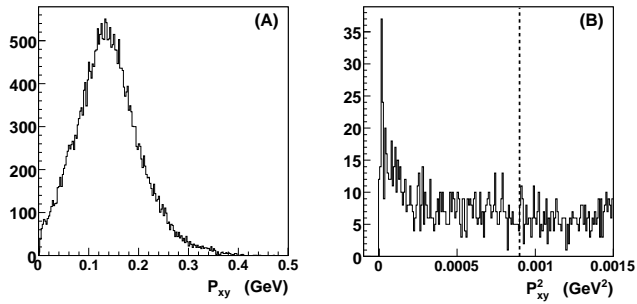,width=\columnwidth}
\caption{(A) Transverse missing 
momentum and (B) transverse missing momentum squared for the reaction 
$\gamma p \to K^+ \Lambda (X)$.  The dashed line shows the cut used at $P^2_{xy}>0.0009$ GeV$^2$.}
\label{PXY}
\end{figure}

A small fraction of the events near zero missing mass in the spectrum of Fig. \ref{MM1} come
from accidentals and double bremsstrahlung.
In the case of double bremsstrahlung it is possible for false EM decay signals caused by 
the reaction $\gamma_1 + \gamma_2p \to K^{+} \Lambda + \gamma_1$ to
mimic the final state of interest $\gamma p \to K^{+} \Lambda \gamma$. 
The $\gamma_1$ from double bremsstrahlung will point down the $z$-axis 
(along the beam), which can also occur if the event is accidental or due to 
inefficiencies in the tagger plane from incorrect electron selection.  
By calculating the transverse missing momentum ($P_{xy}^2=P^2_x+P^2_y$), 
it is possible to eliminate double bremsstrahlung.  
The peak at small values in the distribution in Fig. \ref{PXY} was 
removed by requiring $P^2_{xy}>0.0009$ GeV$^2$ as illustrated by the dashed line.  
Clearly the effect is quite small, however this step is critical for an accurate
measure.

In Fig. \ref{MM1} the tail of the $\pi^{0}$ peak continues into the zero 
missing mass region.  Resolving the two contributions with a simple Gaussian fit does not
give systematically consistent results.  
A technique involving kinematic fitting is required
to separate the background from the events of interest, as well as to separate
the $\pi^{0}$ events from the radiative signal.

\section{Simulations}

A Monte Carlo simulation of the CLAS detector was performed using 
GEANT \cite{geant}, set up for the $g11$ run conditions.
Events were generated for the radiative channel 
($\Sigma^0(1385) \rightarrow \Lambda\gamma$),
the normalization reaction 
($\Sigma^0(1385)\rightarrow\Lambda\pi^0$), 
and several background reactions; see Table \ref{acc2} for a complete list. 
Using the data as a guide, the photon beam energy dependence of 
$K^+$ production and the $K^+$ angular dependence were used iteratively 
to tune the Monte Carlo to match the data.  
After reconstruction, the Monte Carlo momentum distributions for the proton, 
$\pi^-$, and $K^+$ matched (within error bars) to that of the data.  
The generated Monte Carlo events were analyzed using the same analysis
procedure used for the data.

After studying the various channels of interest and background, 
a constant $t$-slope ($d\sigma/d\Omega\sim e^{-bt}$) of $b=$2.0 GeV$^{-2}$ was used for the generated $\gamma p \to
K^+\Lambda(1405)$ channel.
The form of the angular distributions of the cross section from data were used in the generator to
produce all the $\Sigma^{*}$ simulations.

\section{Kinematic Fitting}

The kinematic fitting employed in this analysis technique takes advantage of the information 
in the measured kinematic variables and their uncertainties to fit constraints of energy 
and momentum conservation, thereby improving the measured quantities 
using constraint equations.
This procedure is useful to improve the separation of signal from background.  
The method of Lagrange multipliers is the approach implemented here 
to fit the constraints with a least squares criteria \cite{keller}.

Assume there are $n$ independently measured data values $y$, which in turn
are functions of $m$ unknown variables $q_i$, with $m\leq n$.  
The condition that $y=f_{k}(q_i)$ is introduced, where $f_{k}$ is a 
function dependent on the data points that are being tested for
each $k$ independent variable at each point.

Because each $y_k$ is a measurement with a corresponding standard 
deviation $\sigma_k$, the equation $y_k=f_k(q_i)$ cannot be satisfied 
exactly for $m < n$.  
It is possible to require that the relationship be 
satisfied by defining the $\chi^{2}$ relation such that
\begin{equation}
\chi^{2} = \sum_k { {(y_k - f_k(q))}^2 \over {\sigma_k^2} },
\end{equation}
and require that the preserved values are $q_i$, which are the values of $q$ that minimize $\chi^{2}$.

The unknowns are divided into a set of measured variables ($\vec\eta$), such as 
the measured momentum components, and unmeasured variables ($\vec u$), such as
the missing momentum or the four-vector of an undetected particle in the reaction.  
The variable $\mathcal{L}_i$ is introduced to be used for each 
constraint equation.  The Lagrange multipliers ($\mathcal{L}_i$) are used to write 
the equation for $\chi^{2}$ for a set of constraint equations 
$\mathcal{F}$ such that,
\begin{equation}
\chi^2(\vec\eta,\vec u,\mathcal{L})= (\vec\eta_0-\vec\eta)^T V^{-1} 
	(\vec\eta_0-\vec\eta)+ 2 \mathcal{L}^T  \mathcal{F}(\vec\eta,\vec u), 
\end{equation}
where $\vec\eta_0$ is a vector of initial measured quantities and 
$V^{-1}$ is the inverse of the covariance matrix containing all of the 
resolutions and correlations of the measured variables from the drift
chamber tracking for each charged particle.

The $\chi^{2}$ minimization occurs by differentiating $\chi^{2}$ 
with respect to each of the variables, while linearizing the constraint 
equations and obtaining improved measured values from the fit.  
These values are used as the input 
for a series of iterations.  
The iteration procedure is continued until the difference in magnitude 
between the current $\chi^2$ and the previous value is smaller than 
$\Delta\chi^2_{test}$ ($\le$0.001).
 
The implemented covariance matrix $V$ was corrected 
for multiple scattering and energy loss in the 
target cell, the scattering chamber, and the start counter. 
These corrections to the diagonal terms in the covariance matrix are 
applied according to the distance each charged particle travels
through the corresponding material.  
  
\section{Analysis Procedure}

A useful kinematic fit for a topology that has a particle that is not detected
is the 1C fit.  Such a fit requires that a missing mass hypothesis be used to
constrain the detected four-momentum, leading to three unknowns from the
non-detected particle momentum and four constraints from conservation of
energy and momentum.  To ensure only high quality $\Lambda$ events, an additional
constraint can be implemented on the proton and $\pi^{-}$ tracks to constrain 
the invariant mass to be the known mass of the $\Lambda$.   
After the detected particle tracks are kinematically fit, the events can be 
filtered with a confidence level cut.
In this fit there are three unknowns ($\vec p_{x}$) and five 
constraint equations, four from conservation of energy and momentum and 
the additional invariant mass condition.  No additional constraints are required.
This makes it a 2C kinematic fit. 

To separate the contributions of the 
$\Sigma^{*0}$ EM decay and the strong decay,
the events were fit using the hypotheses for each topology
with the constraint equations,
\begin{equation}
\mathcal{F} = \left[ \begin{array}{c}
	(E_{\pi}+E_{p})^2 -(\vec p_{\pi} + \vec p_p)^2 - M_{\Lambda}^{2} \\
	E_{\text{beam}}+M_p-E_K-E_p-E_{\pi}-E_X \\
	\vec p_{\text{beam}} - \vec p_K - \vec p_p -\vec p_{\pi} -\vec p_X
	\end{array} \right]=\vec 0, 
\end{equation}
where $\vec p_X$ and $E_X$ are the momentum and energy of the 
undetected $\pi^0$ or $\gamma$.

To test the functionality of the kinematic fit
used to separate the radiative signal from the overwhelming $\pi^0$
background, the probability density function \cite{keller}
is used to fit the resulting $\chi^2$ distribution.  The additional constraint on 
the invariant mass of the $\Lambda$ takes the probability density function from
the more difficult to fit one degree of freedom $\chi^2$ distribution (containing a singularity) to the more manageable two degrees of freedom. The fit function takes the form,
\begin{equation}
f(\chi^{2})=\frac{P_{0}}{2}e^{-P_{1}\chi^{2}/2}+P_{2},
\label{fit2df}
\end{equation}
where $P_2$ is a background term, $P_{1}$ is a quantitative closeness 
parameter (which gives a measure of how close the distribution in the 
histogram is to the ideal theoretical $\chi^{2}$ distribution), and 
$P_{0}$ is for normalization.  
For a kinematic fit to a missing $\gamma$ with significant background
contamination from the $\pi^0$, the $\chi^{2}$ distribution will be
highly distorted.
The ideal $P_1$ from a fit to a $\chi^2$ distribution with no background is determined
from simulations.  The deviation of the $P_1$ fit parameter from the
ideal $P_1$ is used as an indicator of the signal to background contribution
going into the kinematic fit under the radiative hypothesis and how
effective a confidence level cut is expected to be for that
given deviation.

\begin{figure}
\epsfig{file=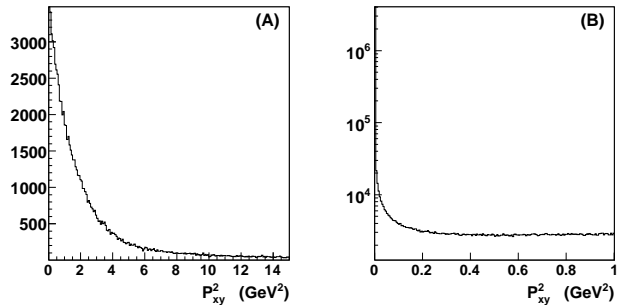,width=\columnwidth}
\caption{ 
(A) $\chi^2$ distribution and (B) confidence level distribution 
for a missing $\pi^{0}$ hypothesis in the kinematic fit.}
\label{conpi0}
\end{figure}

Using the $\pi^{0}$-hypothesis for the kinematic fit, 
the $\chi^{2}$ distribution follows the trend of the
probability density function for two degrees of freedom from Eq. (\ref{fit2df}), 
see Fig. \ref{conpi0}A.
The confidence level in Fig. \ref{conpi0}B is reasonably flat for 
the vast majority of events.  The spike at zero confidence level in Fig. \ref{conpi0}B
is from events that do not satisfy the hypothesis in the kinematic fit.

\begin{figure}
\epsfig{file=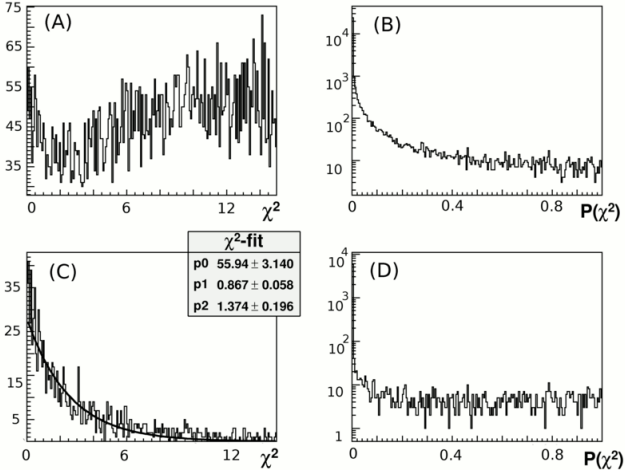,width=\columnwidth}
\caption{(A) The $\chi^2$ distribution and (B) the confidence level 
distribution for a missing $\gamma$ hypothesis in
the kinematic fit before the two-step kinematic fit.  
(C) The $\chi^2$ distribution and the
(D) the confidence level distribution for a missing $\gamma$ hypothesis in
the kinematic fit after the $P^a_{\pi}(\chi^2)<1\%$ cut.
}
\label{cong}
\end{figure}

For the $\gamma$-hypothesis, without any cuts to reduce the $\pi^0$ background, 
the $\chi^{2}$ distribution is not consistent with the expected probability
density function for a 2C fit.  Simulations indicate that the $P_1$ parameter
should be $P_1\sim0.9$, but due to the sensitive nature of the $\chi^{2}$ distribution
for two degrees of freedom, a fit to obtain the $P_1$ parameter does not
return a realistic value.  This can be seen in the distorted shape of the
distribution in Fig. \ref{cong}A.  Additionally, the confidence level 
distribution rises up near the low confidence end (Fig. \ref{cong}B) 
and is clearly not as flat as the distribution
in Fig. \ref{conpi0}B.
This is an indication that the vast majority of data being 
kinematically fit at this stage are not satisfying the base assumption 
of a massless missing particle.  
This suggests that, even with a high confidence level cut,
there is still an overwhelming amount of $\pi^{0}$ events leaking through.  
However, it is possible to take an additional step in the 
kinematic fitting procedure for a cleaner separation.

A two-step kinematic fitting procedure is used to systematically
reduce the large $\pi^0$ background and optimize the extraction of the number of radiative events.  First, a fit to a 
$\pi^{0}$-hypothesis is done and only the low confidence level ($P^a_{\pi}(\chi^2)$)
events are retained, followed by a fit of these candidate events 
to a $\gamma$-hypothesis and retaining the high confidence level ($P^b_{\gamma}(\chi^2)$) events.  
Because of the previous kinematic cuts, there should now be primarily 
a $\pi^{0}$ background and the true EM decay signal.  Any other
background is expected to be very small relative to the radiative signal and will be accounted for
through simulations.  By first fitting to a $\pi^{0}$-hypothesis and 
taking the low confidence level candidates, one reduces the 
probability that the surviving candidates will have a missing mass 
of the $\pi^{0}$ before they are fit to a $\gamma$-hypothesis.

The selection of the confidence level cuts $P^a_{\pi}(\chi^2)$ and $P^b_{\gamma}(\chi^2)$ is derived using
simulations. After testing the ability to recover various mixed ratios on the
order of the expected experimental ratio ($\sim 1\%$),
a Monte Carlo (MC) simulation of the data was studied for a given ratio of the 
$\gamma p \to K^{+} \Sigma^{*0} \to K^{+} \Lambda \pi^{0}$ and 
$\gamma p \to K^{+} \Sigma^{*0} \to K^{+} \Lambda \gamma$ channels. The
optimization occurs when considering both the increase in statistical uncertainty
from a higher $P^a_{\pi}(\chi^2)$ cut and the increase in MC ratio ``recovery'' uncertainty
from a lower $P^a_{\pi}(\chi^2)$ cut.  The ``recovery'' uncertainty is defined by the
difference in the MC generated ratio and the measured ratio or ``recovered'' ratio found by
analyzing the MC with a given $P^a_{\pi}(\chi^2)$ and $P^b_{\gamma}(\chi^2)$.
The final confidence level cut
in $P^b_{\gamma}(\chi^2)$ is determined by the fit parameter $P_1$ indicating
how much $\pi^0$ background is left after the $P^a_{\pi}(\chi^2)$ cut.  Again,
the statistical uncertainty and the MC ratio ``recovery'' uncertainty are considered
in the optimization of the $P^b_{\gamma}(\chi^2)$ cut.

The results of the optimization study indicate that a confidence level 
cut of $P^a_{\pi}(\chi^2)<1\%$ sufficiently reduces the $\pi^0$ background 
so that a $P^b_{\gamma}(\chi^2)>10\%$ cut
can be used to isolate the radiative signal in the kinematic fit to $\gamma$.

After the two-step kinematic fitting procedure, one can again study the 
$\gamma$-hypothesis $\chi^{2}$ fit.  It now looks more like a 
standard distribution for two degrees of freedom, returning a value of
$P_1=0.87\pm0.06$, see Fig. \ref{cong}C.
The confidence level now appears relatively flat in 
Fig. \ref{cong}D, as it should. 
This is an indication that an improvement has been made on the quality 
of candidates going into the fit with respect to the hypothesis.  
This gives some assurance that the candidates going into the 
secondary fit can be accurately filtered with a confidence level cut.

To ensure the quality of the $\pi^0$ extraction, 
the same two-step kinematic fitting 
procedure is done by first fitting to a $\gamma$ hypothesis and taking the
low confidence level $P^a_{\gamma}(\chi^2)$ candidates, then fitting to the $\pi^0$ hypothesis and
taking only the high confidence level $P^b_{\pi}(\chi^2)$ candidates.

\begin{figure}
\epsfig{file=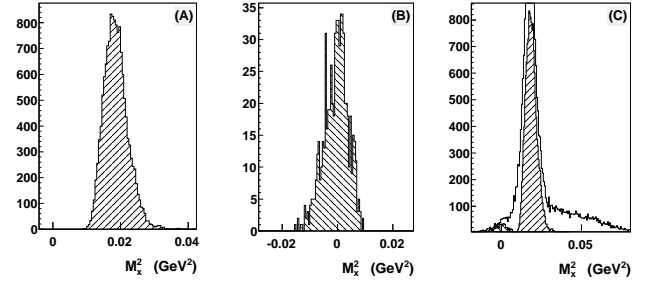,width=\columnwidth}
\caption{(A) The $n_{\pi}$ counts extracted using the confidence level cuts $P^a_{\gamma}<1\%$ and $P^b_{\pi}>10\%$.
(B) The $n_{\gamma}$ counts extracted using the confidence level cuts $P^a_{\pi}<1\%$ and $P^b_{\gamma}>10\%$.
(C) The counts $n_{\pi}$ and $n_{\gamma}$ shown in the spectrum before any kinematic fit.}
\label{spectrum}
\end{figure}

Once the confidence level cuts are optimized for extracting both the $\pi^0$ and
radiative signal, the final selected candidates for each case can be seen in the 
missing mass spectrum, see Fig. \ref{spectrum}.  The extracted counts are shown for
(A) the $\pi^0$, (B) the electromagnetic signal, and (C) together in the full spectrum of
the missing mass squared.  The final raw yields taken directly from the kinematic fit are $n_{\gamma}=635$ and $n_{\pi}=13950$.

\begin{table*}
\caption{Acceptances (in units of $10^{-3}$) for the channels used in the 
calculation of the branching ratio. 
Here there is a $P^a(\chi^2)<1\%$ confidence level used with a 
$P^b(\chi^2)>10\%$ cuts.  The uncertainties listed are statistical only.}
\begin{center}
\begin{tabular}{lcccc}
Reaction & $A_\pi$ & $A_\gamma$ & $A_{\gamma\pi}$ \\ \hline
$\Lambda(1405)\rightarrow\Sigma^0\pi^0$
       &0.0495$\pm$0.0031 &0.001$\pm$0.0001 & 1.189$\pm$0.019 \\
$\Lambda(1405)\rightarrow\Sigma^+\pi^-$
       & 0.029$\pm$0.002 & 0.0013$\pm$0.0001 & 0.0078$\pm$0.001   \\   
$\Lambda(1405)\rightarrow\Lambda\gamma$
       & 0.0011$\pm$0.0001 &1.65$\pm$0.031& 0.0223$\pm$0.002 \\
$\Lambda(1405)\rightarrow\Sigma^0\gamma$
       & 0.170$\pm$0.012 & 0.191$\pm$0.009 & 0.437$\pm$0.013 \\
$\Sigma(1385)\rightarrow\Lambda\pi$    
       & 1.421$\pm$0.0278 & 0.0321$\pm$0.002 & 0.0312$\pm$0.002 \\
$\Sigma(1385)\rightarrow\Sigma^+\pi^-$ 
       & 0.161$\pm$0.01 & 0.00254$\pm$0.001& 0.00138$\pm$0.0006 \\
$\Sigma(1385)\rightarrow\Lambda\gamma$ 
       & 0.0184$\pm$0.002 &  2.335$\pm$0.039 &  0.0704$\pm$0.005 \\
$\Sigma(1385)\rightarrow\Sigma^0\gamma$
       &  0.191$\pm$0.011 & 0.058$\pm$0.0001 & 0.225$\pm$0.015 \\ 
$\Lambda K^{*+}\rightarrow K^{+}\pi^{0}$
       &  0.213$\pm$0.010 & 0.010$\pm$0.006  & 2.931$\pm$0.051 \\        
$\Lambda K^{*+}\rightarrow K^{+}\gamma$
       &  0.0022$\pm$0.0001 & 0.158$\pm$0.003 & 2.351$\pm$0.046 \\          
       \hline
\end{tabular}
\end{center}
\label{acc2}
\end{table*}

The $\pi^0$ leakage into the $\gamma$ channel is the dominant 
correction to the branching ratio. 
The final result also needs to be corrected for backgrounds, 
such as $K^* \to K^+X$ and decays to $\Sigma^+\pi^-$, 
as well as the contributions from 
$\Lambda(1405) \to \Lambda\gamma$.  
Taking these backgrounds into consideration, and 
following the notation of Taylor \etal~\cite{Taylor}, 
the branching ratio $R=N(\Lambda \gamma)/N(\Lambda \pi)$ is
\begin{eqnarray}
R&=&\frac{1}{\Delta n_\pi A^\Sigma_\gamma
(\Lambda\gamma)
- \Delta n_\gamma A^\Sigma_\pi(\Lambda\gamma)} \nonumber \\
& & \times
\left[\Delta n_\gamma
 \left(A^\Sigma_\pi(\Lambda\pi)+\frac{R^{\Sigma\pi}_{\Lambda\pi}}{2}
A^\Sigma_\pi(\Sigma\pi)\right)\right. \nonumber \\
& &\left. -\Delta n_\pi
 \left(A^\Sigma_\gamma(\Lambda\pi)+\frac{R^{\Sigma\pi}_{\Lambda\pi}}{2}
A^\Sigma_\gamma(\Sigma\pi)\right)\right],
\label{finalR}
\end{eqnarray}
where terms starting with $A$ are acceptance factors (given below) and 
\begin{eqnarray}
\Delta n_\pi&=&n_\pi-N_\pi(\Lambda^*\rightarrow \Sigma^+\pi^-)
                   -N_\pi(\Lambda^*\rightarrow \Sigma^0\pi^0) \nonumber \\
       & &         -N_\pi(\Lambda^*\rightarrow \Sigma^0\gamma)  
                   -N_\pi(\Lambda^*\rightarrow \Lambda \gamma)\nonumber\\
       & &	   -N_\pi(K^*\rightarrow K \pi^0), \\
\Delta n_\gamma&=&n_\gamma-N_\gamma(\Lambda^*\rightarrow \Sigma^+\pi^-)
                   -N_\gamma(\Lambda^*\rightarrow \Sigma^0\pi^0) \nonumber\\
        & &        -N_\gamma(\Lambda^*\rightarrow \Sigma^0\gamma)
		   -N_\gamma(\Lambda^*\rightarrow \Lambda \gamma)\nonumber\\
       & &	   -N_\gamma(K^*\rightarrow K \gamma),
\end{eqnarray}
with $n_\gamma$ ($n_\pi$) equal to the yield of the kinematic fits, 
representing the measured number of photon (pion) candidates.  
In the notation used, lower case $n$ represents an observed number of counts, 
while upper case $N$ represents the acceptance corrected or derived 
quantities.  Here the $\pi$ and $\gamma$ subscripts indicate the 
kinematic fit hypothesis and the decay channel is shown in 
parentheses (note that $\Lambda^*$ denotes the $\Lambda(1405)$).  
These corrections are necessary to take into account due to the fact 
that the background underneath the $\Sigma(1385)$ is not 
zero, which could lead to an over-counting of the $\Sigma(1385)$ contribution.  
For the detector acceptance, the notation has the 
pion (photon) hypothesis from the $\Sigma(1385)$ decay given by 
$A^\Sigma_{\pi}$ ($A^\Sigma_{\gamma}$), so that 
$A^\Sigma_\gamma(\Lambda\pi)$ denotes the relative leakage of 
the $\Sigma^{*0} \to \Lambda\pi$ decay channel into the $\Lambda\gamma$ 
extraction 
and $A^\Sigma_\pi(\Lambda\gamma)$ denotes the relative leakage of the 
$\Lambda\gamma$ decay channel into the $\Lambda\pi$ extraction.
The form of the ratio given in Eq. (7) is developed in more detail in the Appendix.

\section{Background Contributions}

Table \ref{acc2} lists all decay channels taken into consideration and 
the value of the acceptance for the confidence level cuts 
$P^a_{\pi}(\chi^{2}) < 1$\% followed by 
$P^b_{\gamma} (\chi^{2})   >  10$\% for the $\gamma$-hypothesis and 
$P^a_{\gamma} (\chi^{2}) < 1$\% followed by 
$P^b_{\pi}(\chi^{2})   >  10$\% for the $\pi^0$-hypothesis.
To use these acceptance terms to correct the signal yields, an estimate of the number 
$n_\Lambda$ for the $\Lambda(1405)$ in the event sample is required.  
The corrections for the $\gamma$ channel are given by
\begin{eqnarray}
N_{\gamma}(\Lambda^*\rightarrow\Lambda\gamma)&=&
 { A^\Lambda_{\gamma}(\Lambda\gamma) BR(\Lambda^*\rightarrow\Lambda\gamma)
 n_\Lambda \over A^\Lambda_{\gamma\pi}(\Sigma^0\pi^0)
 +A^\Lambda_{\gamma\pi} (\Sigma^+\pi^-)},\\
N_{\gamma}(\Lambda^*\rightarrow\Sigma^0\gamma)&=&
 { A^\Lambda_{\gamma}(\Sigma^0\gamma) BR(\Lambda^*\rightarrow\Sigma^0\gamma)
 n_\Lambda \over A^\Lambda_{\gamma\pi}(\Sigma^0\pi^0)
 +A^\Lambda_{\gamma\pi} (\Sigma^+\pi^-)},\\
N_{\gamma}(\Lambda^*\rightarrow\Sigma^0\pi^0)&=&
{ A^\Lambda_{\gamma}(\Sigma^0\pi^0) n_\Lambda
 \over A^\Lambda_{\gamma\pi}(\Sigma^0\pi^0)
 +A^\Lambda_{\gamma\pi} (\Sigma^+\pi^-)},\\
N_{\gamma}(\Lambda^*\rightarrow\Sigma^+\pi^-)&=&
{ A^\Lambda_{\gamma}(\Sigma^+\pi^-) n_\Lambda
 \over A^\Lambda_{\gamma\pi}(\Sigma^0\pi^0)
 +A^\Lambda_{\gamma\pi} (\Sigma^+\pi^-)},
\end{eqnarray}
where $BR$ is the branching ratio for the decay shown, 
and likewise for the $\pi^{0}$ channel. 

Isospin symmetry is assumed so 
that $BR(\Sigma^0\pi^0)=BR(\Sigma^+\pi^-)=BR(\Sigma^-\pi^+)\approx 1/3$ 
for the $\Lambda(1405)$ decay channels.  
The subscript ``$\gamma\pi$'' denotes the acceptance for events that do 
not satisfy the confidence level cuts for either hypotheses of the kinematic 
fit ({\it i.e.} it is likely to come from some background reaction). 
The values for $BR(\Lambda(1405) \to \Lambda \gamma)=5.4\pm0.2\times 10^{-4}\%$ and
$BR(\Lambda(1405) \to \Sigma^0 \gamma) = 2.0\pm0.1\times 10^{-4}\%$ 
are taken from Ref. \cite{Burkhardt}.

Contributions from the $\Sigma\pi$ decay of the $\Sigma^0(1385)$ are also considered.  The term
in Eq. (\ref{finalR}) that takes the $\Sigma^*\to \Sigma^+\pi^-$ counts into consideration
uses $R^{\Sigma\pi}_{\Lambda\pi}= 0.135\pm0.011$ \cite{PDG}, with the acceptance for the individual
channels subject to the radiative ($\pi^0$) hypothesis and denoted as $A^{\Sigma}_{\gamma}(\Sigma\pi)$ 
($A^{\Sigma}_{\pi}(\Sigma\pi)$).  The $\Sigma^*\to \Sigma^0\gamma$ decay branch is also considered, however
no measured branching ratio currently exists for this channel.  The acceptances are very small and
using the higher theoretical prediction of the Algebraic model \cite{Bijker} yielded only negligible  
contributions to the background.

\begin{figure}
\epsfig{file=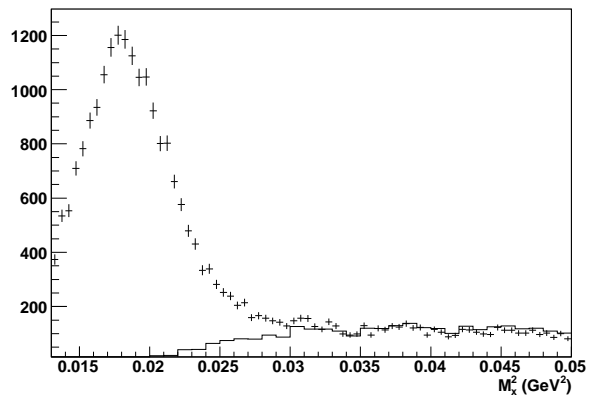,width=\columnwidth}
\caption{Missing mass of $\gamma p \to K^+ \Lambda (X)$ for 
data (points with error bars) and 
Monte Carlo simulations for the $\gamma p \to K^+ \Lambda(1405)$ 
reaction (histogram) normalized to the data.}
\label{bgfill}
\end{figure}

In order to find $n_\Lambda$, one can look at the events 
for which neither the $\gamma$ nor the $\pi^0$ hypothesis is satisfied.  
The value of $n_\Lambda$ is difficult to determine due to the 
non-Breit-Wigner shape of the $\Lambda(1405)$ decay.  
A better approach is to use a Monte Carlo simulation to fill the background according
to its internal decay kinematics and normalize it to the data such that 
the MC matches the data, thereby giving an estimate of $n_\Lambda$.  
Figure \ref{bgfill} shows the MC simulation normalized to the data, 
giving the estimate used for $n_\Lambda$.  
This can be used to correct all backgrounds except for the $K^{*}$.  The final estimate
found is $n_\Lambda=4085$.

\begin{figure}
\epsfig{file=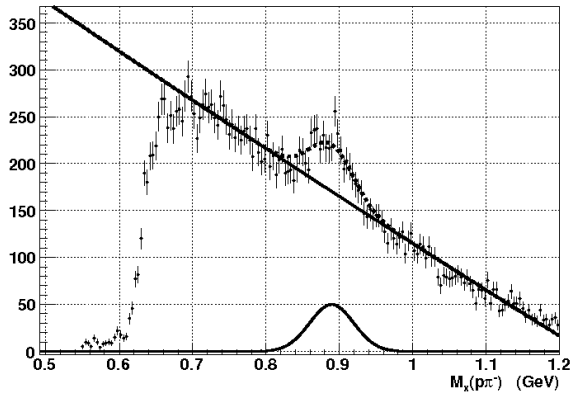,width=\columnwidth}
\caption{Missing mass off the $\Lambda$ fit with a Gaussian for the
mass off of the $K^{+}$ in the window of 1.35-1.5 GeV.}
\label{mc_fit_bg1}
\end{figure}

The $\gamma p \to K^{*0} \Sigma^{+}$ reaction was investigated with 
the MC simulation and compared with data. 
This background was determined to have a negligible effect on the 
final result, since there is no $\Lambda$ in the final state. 
For the $\gamma p \to K^{*+}\Lambda$ reaction, few events survive all 
of the cuts. 
To include corrections for the few events that do survive, an estimate 
of the $K^{*+}$ background must be established.  
The correction for this background has the form 
\begin{eqnarray}
N_\pi(K^*\rightarrow K^{+} \pi^0) = &\\ 
\frac{A_{\pi^0}(K^{*+}\to K^{+}\pi^0) n(K^{*+} \to K^{+}\pi^{0})}{A_{\pi} (K^{*+} \to K^{+}\pi^{0})},&\nonumber
\label{kcal}
\end{eqnarray}
where $A_{\pi}(K^{*+}\to K^{+}\pi^0)$ is the acceptance for the $K^{*+}\to K^{+}\pi^{0}$
channel under the $\pi^0$-hypothesis while $A(K^{*+}\to K^{+}\pi^0)$ is the
acceptance of the $K^{*+}\to K^+\pi^0$ channel which is dependent on the
extraction method to obtain the $K^{*+}\to K^{+}\pi^{0}$ counts.
$n(K^{*+} \to K^{+}\pi^{0})$ is the estimated number of 
$K^{*+} \to K^{+}\pi^{0}$ events in the data sample.   
Similarly, the radiative decay of the $K^{*}$ has the form
\begin{eqnarray}
N_\gamma(K^*\rightarrow K \gamma) = \frac{3}{2}R(K^{*+} \to K^{+}\gamma)&\\
\times \frac{A_{\gamma}(K^{*+} \to K^{+}\gamma)}{A(K^{*+} \to K^{+}\pi^0)}n(K^{*+} \rightarrow K^+\pi^{0}),&  
\nonumber
\end{eqnarray}
with $N_\pi(K^* \rightarrow K \pi^0)$ from Eq. (\ref{kcal})  
and $BR(K^{*+} \to K^{+}\gamma) \simeq 9.9 \times 10 ^{-4}$.
An estimate of the number of $K^{*}$ events was obtained 
by matching the MC simulations to the data.
The $K^{*+} \to K^{+}\pi^{0}$ mass distribution has been fit
as shown in Fig. \ref{mc_fit_bg1}.
In addition a fit is more easily obtained over
the range that allows the higher part of the excited state mass spectrum
to pass through.  Using the resulting Gaussian fit of the $K^*$ peak while studying the
mass off of the $K^{+}$ for mass
windows ranging from 1.34-1.5 GeV to 1.34-1.8 GeV, we are able to extrapolate down to
the nominal $\Sigma^{*0}$ mass cut (1.34-1.43 GeV), see Fig. \ref{offKp}.
Both methods gave similar results for $n(K^{*+} \to K^+\pi^0)$ used for the 
background correction in the final ratio.  The extrapolated number of
$K^* \to K^{+}\pi^0$ events present were 1207,
14.8 of which passed the $\pi^0$-hypothesis.

Each background contribution is enumerated in Table \ref{background}.  The table
gives the final breakdown of statistics for each term used in Eqs. (8)-(9) to achieve 
the corrected counts.

\begin{table}
\caption{Breakdown of statistics for each term in Eqs. (8)-(9) for the $\Lambda(\gamma)$ and $\Lambda(\pi^0)$
hypotheses. Each listed channel is subtracted from the raw counts found from the
kinematic fit in each case.  The uncertainties are statistical only.  The $\Sigma(1385)\rightarrow\Sigma^+\pi^-$
contributions are of the order of $10^{-5}$ and smaller and are not included in the table.}
\begin{center}
\begin{tabular}{lccc}
Reaction & $\Lambda(\gamma)$ & $\Lambda(\pi^0)$  \\ \hline
Raw counts & 635 & 13950  \\ \hline
$\Lambda(1405)\rightarrow\Sigma^0\pi^0$
       &3.41$\pm$0.36 &168.94$\pm$11.65\\
$\Lambda(1405)\rightarrow\Sigma^+\pi^-$
       & 4.44$\pm$1.09 & 98.98$\pm$7.80\\   
$\Lambda(1405)\rightarrow\Lambda\gamma$
       & 3.04$\pm$0.59 &0.01$\pm$0.00 \\
$\Lambda(1405)\rightarrow\Sigma^0\gamma$
       & 0.13$\pm$0.04 & 0.12$\pm$0.04 \\
$\Lambda K^{*+}\rightarrow K^{+}\pi^{0}$
       &  0.00$\pm$0.00 & 27.41$\pm$1.71\\        
$\Lambda K^{*+}\rightarrow K^{+}\gamma$
       &  0.03$\pm$0.00 & 0.00$\pm$0.00\\ \hline
Corrected counts & 623.92$\pm$25.23 & 13654.56$\pm$118.95  \\ \hline       
\end{tabular}
\end{center}
\label{background}
\end{table}

\subsection{RESULTS}
The resulting corrected counts $\Delta n_{\gamma}=623.92\pm25.23$ and $\Delta n_{\pi}=13654.56\pm118.95$
are used in Eq. (\ref{finalR}) to obtain a ratio of 1.42 $\pm0.12(\text{stat})\%$.
The value of each cut was varied to study the effect on the final acceptance corrected ratio.  
For each variation the new acceptance terms along with the background contributions in Eq. (\ref{finalR}) were also
recalculated.  Each major systematic uncertainty contribution is listed in 
Table \ref{systematic}, as described in detail below.

Particle identification was done by calculating the velocity for 
a particle of a given mass, using the measured momentum, and 
compared with that expected from the measured path length and 
time-of-flight.  Kaons, protons and pions are selected based on 
the difference of the calculated and measured velocity, 
called $\Delta\beta$.  Variations of the width of the $\Delta
\beta$ cut to identify particles gave slightly different values 
of the final ratio, shown in line (1) of Table \ref{systematic}.

The distance of closest approach (DOCA) cut for the proton and 
$\pi^-$, used to reconstruct the $\Lambda$ momentum, 
was varied and the stability of the final ratio was examined. 
In the stable region, corresponding to cuts in the range from 3 cm to 14 cm, the
effect on the final ratio is mostly toward higher values, for a 
larger DOCA cut value, as listed in line (2). 

Similarly, the value of the cut on the transverse momentum, 
$P_{xy}$, was varied. The ratio stabilizes starting at the cut point shown in Fig. 
\ref{PXY}.  A series of cuts were used starting at 0.0009 GeV$^2$ and ending at 0.0025 GeV$^2$
to study the effect on the final ratio, given by line (3).

The Monte Carlo simulations for various background reactions 
were done assuming a $t$-dependent slope of 2.0 GeV$^{-2}$ for the 
differential cross sections, based on Regge theory, as described 
earlier.  The value of the $t$-slope is not known precisely, and 
was varied by $\pm$25\%.  The effect on the final 
ratio is shown in line (4).

The number of counts for the $\Lambda(1405)$ and $K^*$ 
backgrounds were determined from fits to the data, using 
comparisons of shapes from Monte Carlo simulations with the data 
shown in Figs. \ref{bgfill} and \ref{mc_fit_bg1}.  The 
uncertainty in the number of counts for these backgrounds also 
affects the final ratio, as shown in lines (5) and (6).

To look at the systematic dependence on the choice of the 
confidence level cuts, the range defined by the uncertainty for 
$P_{\pi}^{b}(\chi^{2})$ was checked.  As previously 
described, the Monte Carlo studies lead to the set of optimal 
$P^{a}_{\pi}$ cuts for a given $P^{b}_{\gamma}$.  These 
optimal cuts allow one to recover, in our standard analysis 
framework, the ratio of $\Lambda\gamma$ to $\Lambda\pi^0$ decay that was used as
input into the Monte Carlo simulations. To maximize counting statistics while 
minimize this uncertainty, the optimal cuts chosen for the 
analysis were $P^{b}_{\pi}<1\%$ and $P^{b}_{\gamma}>10\%$.

The variation in the branching ratio is studied by selecting the 
confidence level cuts that lie slightly outside the optimization 
region found in simulations.  In this way the largest range for 
$P^{b}_{\gamma}$ and $P^{a}_{\pi}$ can be tested while still 
respecting the cuts derived from the optimization map. Using the 
full range of ratios in Table \ref{con_cut2}, the largest and 
smallest values show the variation for different choices of 
confidence level cuts, and is listed in line (7).

The branching ratio of the radiative decay of the 
$\Lambda(1405)$ affects the final result.  Because this value is 
not measured directly, but is taken from the calculated value in Ref. \cite{Burkhardt}, there is some uncertainty
in it. Using the range of values for this branching ratio given in the literature, and recalculating its effect on our result, leads to 
the uncertainty quoted in line (8).

Table \ref{systematic} shows a summary of the systematic studies 
and the higher and lower value of the extracted ratio based on the 
variations mentioned for each type of uncertainty.
The deviation of the ratio is defined by the difference from the 
quoted ratio of $R=1.42\%$.  These deviations, shown in columns 
3 and 5 of  Table \ref{systematic}, are added in quadrature to 
give the total systematic uncertainty.

\begin{table*}[htb]
\caption{Ranges of systematic variation of the final ratio, 
given in percent, along with the deviation from the central 
value.}
\begin{center}
\begin{tabular}{lccccc}
Source & Low Value & Low Deviation & High Value & High Deviation \\ \hline
(1)Particle identification & 1.380 & $-0.040$ & 1.490 & +0.070   \\
(2)$p\pi^-$ DOCA cut point & 1.350 & $-0.007$ & 1.480 & +0.060   \\
(3)Transverse momentum $P_{xy}$ & 1.415 & $-0.005$ & 1.433 & +0.013 \\
(4)Monte Carlo $t$-dependence &1.380 & $-0.040$ & 1.440 & +0.020  \\
(5)$\Lambda(1405)$ counts & 1.420  & $-0.000$ & 1.470 & +0.050  \\
(6)$K^{*}$ counts & 1.420 & $-0.000$ & 1.431 & +0.011 \\
(7)$P(\chi^{2})$ cut points & 1.388 & $-0.032$ & 1.448 & +0.028 \\
(8)$\Lambda(1405)\to\Lambda\gamma$ correction & 1.390 & $-0.030$ & 1.420 & +0.000 \\ \hline
Total Uncertainty	&    & $-0.072$	&	&  +0.112 \\ \hline
\end{tabular}
\end{center}
\label{systematic}
\end{table*}

\begin{table}
\caption{Dependence of corrected branching ratio for variation of 
the confidence level cuts shown.}
\begin{center}
\begin{tabular}{lccc}
$P^{b}_{\gamma}(\%)$ & $P^{a}_{\pi}(\%)$ & R$(\%)$ & \\ \hline
15 & 7.5 & 1.388$\pm$ 0.12 &  \\
15 & 5 & 1.390$\pm$ 0.12 &  \\  
10 & 5 & 1.422$\pm$ 0.12 & \\
10 & 1 & 1.420$\pm$ 0.12 &  \\
10 & 0.5 & 1.421$\pm$ 0.12 &  \\   
5 & 0.1 & 1.448$\pm$ 0.12 &  \\
5 & 0.05 & 1.436$\pm$ 0.12 &  \\  \hline 
\end{tabular}
\end{center}
\label{con_cut2}
\end{table}

The range of the systematic uncertainty for $R$ in Table \ref{systematic}
is smaller than the statistical uncertainty, in part because each 
combination of cuts has a large overlap of events 
({\it i.e.} the same subset of events is present for all choices of cuts).
Since the kinematic fit requires a constraint on the $\Lambda$ mass, 
the kinematic cut on the invariant mass of the $p\pi^-$ has no effect. 

The final calculated ratio, given in percent, is
\begin{eqnarray}
R^{\Lambda \gamma}_{\Lambda\pi}=&\frac{\Gamma[\Sigma^0(1385)\rightarrow\Lambda\gamma]}{\Gamma[\Sigma^0(1385)\rightarrow\Lambda\pi^0]}\nonumber\\
=& 1.42\pm0.12(\text{stat})_{-0.07}^{+0.11}(\text{sys})\%.
\label{ratt}
\end{eqnarray} 

Previously published work \cite{Taylor} on this branching ratio 
yielded a ratio of $1.53\pm0.39^{+0.15}_{-0.17}$ \%.  
The value given here is consistent within uncertainties of the 
previous value, but has smaller uncertainties.
The smaller uncertainty is important, as the previous uncertainty 
was on the same order as the theoretical meson cloud corrections 
to the EM decay of the $\Delta$.  If similar meson cloud 
corrections are to be proven true for EM decay of the 
$\Sigma^{*0}$ baryon, then the smaller experimental uncertainty 
is a significant improvement.

The width for the branching ratio achieved comes from the use of the full width of the $\Sigma^{*0}$,
which is $\Gamma(\Sigma^{*0})_{\text{Full}}=36.0\pm5.0$ MeV with the branching ratio that the radiative signal is
being normalized to, which is the $R(\Sigma^{*0}\to\Lambda\pi^0)=87.0\pm1.5\%$ \cite{PDG}.  The partial width
calculation is then
\begin{eqnarray}
\Gamma_{\Sigma^{*0}\to \Lambda\gamma}=&R^{\Lambda\gamma}_{\Lambda\pi}
R(\Sigma^{*0}\to\Lambda\pi^0)\Gamma(\Sigma^{*0})_{\text{Full}}& \nonumber\\
=&445\pm80~\text{keV},&
\label{width}
\end{eqnarray}
where a systematic uncertanty in $R^{\Lambda\gamma}_{\Lambda\pi}$ of $\pm0.11\%$ is use in combination with the
statistical uncertainty.  Note that a large part of the uncertainty in Eq.
(\ref{width}) comes from the uncertainty of the full width $\Gamma(\Sigma^{*0})_{\text{Full}}$.

These results verify that the partial width is indeed significantly larger than leading
theoretical predictions, indicating that meson cloud effects are an important consideration
for future calculations.

The radiative decay $\Sigma^{*0}\to\Lambda\gamma$ is made up of M1 and E2 electromagnetic transitions.
Assuming that the E2 amplitude is very small, one can calculate the transition 
magnetic moment from the measured radiative width \cite{Landsberg},
\begin{eqnarray}
\mu_{\Sigma^{*0}\to \Lambda\gamma}=\sqrt{\frac{2M^2_p\Gamma_{\Sigma^{*0}\to \Lambda\gamma}}{\alpha
p^3_{\gamma}}}\mu_N=2.75\pm0.25\mu_N,
\end{eqnarray}
where $p_{\gamma}$ is the photon momentum, $M_p$ is the mass of the proton, $\alpha=e^2/4\pi\sim1/137$, and
$\mu_N$ is the nuclear magneton.  The value for the transition magnetic moment is larger than most
model predictions even within the experimental uncertainty.  For example the naive quark model predicts
$\mu_{\Sigma^{*0}\to \Lambda\gamma}=2.28\mu_N$ \cite{Dhir}. 
We hope that this measurement, along with others to come, will motivate 
theorists to understand the effect of the meson cloud on the magnetic 
moment and hence extend our knowledge of the quark wavefunctions in 
the decuplet baryons.

\section{Acknowledgment}
The authors thank the staff of the Thomas
Jefferson National Accelerator Facility who made this experiment possible.
This work was supported in part by 
the Chilean Comisi\'on Nacional de Investigaci\'on Cient\'ifica y Tecnol\'ogica (CONICYT),
the Italian Istituto Nazionale di Fisica Nucleare,
the French Centre National de la Recherche Scientifique,
the French Commissariat \`{a} l'Energie Atomique,
the U.S. Department of Energy,
the National Science Foundation,
the UK Science and Technology Facilities Council (STFC),
the Scottish Universities Physics Alliance (SUPA),
and the National Research Foundation of Korea.
The Southeastern Universities Research Association (SURA) operates the
Thomas Jefferson National Accelerator Facility for the United States
Department of Energy under contract DE-AC05-84ER40150.

\appendix

\subsection{Appendix: Ratio Derivation}
To calculate the
ratio in Eq. (\ref{finalR}), the leakage of the  $\pi^0$ region into the $\gamma$ region (and vice-versa) is the dominant correction.
Taking just these two channels into consideration, the number of 
{\it true} counts can be represented as $N(\Lambda \gamma)$ for the $\Sigma^{*0} \to \Lambda \gamma$ channel and
$N(\Lambda \pi)$ for the $\Sigma^{*0} \to \Lambda \pi^{0}$ channel.  
The acceptance under the $\Sigma^{*0} \to \Lambda \gamma$
hypothesis can be written as $A_\gamma(X)$, with the subscript showing 
the kinematic fit hypothesis type and, in parentheses, 
the channel used in the Monte Carlo for the acceptance.  
For example, the calculated acceptance for the $\Sigma^{*0} \to \Lambda \gamma$
channel under the $\Sigma^{*0} \to \Lambda \gamma$ hypothesis is 
$A_\gamma(\Lambda \gamma)$, whereas under the $\Sigma^{*0} \to \Lambda \pi^{0}$ 
hypothesis it is $A_\pi(\Lambda \gamma)$.  
It is now possible to express the {\it measured} values 
$n_{\gamma}$ and $n_{\pi}$ as
\begin{equation}
n_{\gamma}=A_{\gamma}(\Lambda\gamma)N(\Lambda\gamma)+A_{\gamma}(\Lambda \pi)N(\Lambda \pi)
\label{eq_1}
\end{equation}
\begin{equation}
n_{\pi}=A_{\pi}(\Lambda\pi)N(\Lambda\pi)+A_{\pi}(\Lambda \gamma)N(\Lambda \gamma). \\
\label{eq_2}
\end{equation}
The desired branching ratio of the radiative channel to the $\pi^{0}$ channel 
using the {\it true} counts is then $R=N(\Lambda \gamma)/N(\Lambda \pi)$.
This can be obtained by dividing Eq. (\ref{eq_1}) by Eq. (\ref{eq_2}) 
expressed in terms of $R$ as
\begin{equation}
\frac{n_{\gamma}}{n_{\pi}} = \frac{RA_{\gamma}(\Lambda \gamma)+A_{\gamma}(\Lambda \pi)}{A_{\pi}(\Lambda\pi)+RA_{\pi}(\Lambda \gamma)},
\label{eq_3}
\end{equation}
then solving for $R$. Expressed in terms of measured values and acceptances,
the branching ratio is
\begin{equation}
R = \frac{n_{\gamma}A_{\pi}(\Lambda\pi)-n_{\pi}A_{\gamma}(\Lambda \pi)}{n_{\pi}A_{\gamma}(\Lambda\gamma)-n_{\gamma}A_{\pi}(\Lambda \gamma)}.
\label{eq_4}
\end{equation}

Equation (\ref{eq_4}) uses the assumption that contributions from the
$\Sigma(1385)$ will only show up as $\Lambda \gamma$ or $\Lambda \pi^{0}$, 
neglecting the $\Sigma(1385)\to \Sigma\pi$ channel.  
An estimate of the total number of $\Sigma(1385)$'s produced using the 
$\Lambda\pi^{0}$ channel is
\begin{equation}
N(\Sigma^{*0})=\frac{N(\Sigma^{*} \to \Lambda \pi^{0})}{R(\Sigma^{*}\to \Lambda \pi^{0})A(\Sigma^{*}\to\Lambda\pi^{0})},
\label{eq_5}
\end{equation}
where $R(\Sigma^{*}\to\Lambda\pi^{0})$ is the branching ratio of the 
$\Sigma(1385)$ decay to $\Lambda \pi^{0}$ and
$A(\Sigma^{*}\to\Lambda \pi^{0})$ is the acceptance for that channel.  
An estimate of the number of 
$\Sigma(1385)\to\Sigma^{+}\pi^{-} \to p\pi^0\pi^-$ counts that would 
contribute to the $\pi^{0}$ peak is then:
\begin{eqnarray}
&N(\Sigma^{*}\to \Sigma^{+} \pi^{-})&\nonumber\\
=&R(\Sigma^{*}\to \Sigma^{+} \pi^{-})A(\Sigma^{*}\to \Sigma^{+} \pi^{-})N(\Sigma^{*0})&\nonumber\\
=&\frac{R(\Sigma^{*}\to \Sigma^{+} \pi^{-})A(\Sigma^{*}\to \Sigma^{+} \pi^{-})}{R(\Sigma^{*}\to
\Lambda \pi^{0})A(\Sigma^{*}\to \Lambda \pi^{0})}N(\Sigma^{*}\to \Lambda \pi^{0})&,
\label{eq_6}
\end{eqnarray}
where $R(\Sigma^{*}\to\Sigma^{+}\pi^{-})$ is the branching ratio of the 
$\Sigma(1385)$ to decay into $\Sigma^{+}\pi^{-}$ and 
$A(\Sigma^{*}\to\Sigma^{+}\pi^{-})$ is
the corresponding acceptance after all cuts.  
It is possible to simplify the expression by using,
$$
R^{\Sigma\pi}_{\Lambda\pi}= \frac{R(\Sigma^{*}\to \Sigma^{\pm} \pi^{\mp})}{R(\Sigma^{*}\to \Lambda \pi^{0})}=0.135\pm0.011,
$$
using the PDG average value \cite{PDG}.  
The two charged combinations of the $\Sigma\pi$ decay have equal probability.  
The Clebsch-Gordon coefficient for the $\Sigma^{*}\to\Sigma^0\pi^0$ 
decay is zero, assuming isospin symmetry.
The observed counts, expressed in terms of true counts and corresponding 
acceptances for each hypothesis, becomes
\begin{equation}
n_{\gamma}=A_{\gamma}(\Lambda\gamma)N(\Lambda\gamma)+(A_{\gamma}(\Lambda \pi)+\frac{R^{\Sigma\pi}_{\Lambda\pi}}{2}A_{\gamma}(\Sigma \pi))N(\Lambda \pi)
\label{eq_7}
\end{equation}
and
\begin{equation}
n_{\pi}=(A_{\pi}(\Lambda\pi)+\frac{R^{\Sigma\pi}_{\Lambda\pi}}{2}A_{\pi}(\Sigma \pi))N(\Lambda\pi)+(A_{\pi}(\Lambda \gamma))N(\Lambda \gamma).
\label{eq_8}
\end{equation}
Solving for $R$ will result in a branching ratio that includes all needed 
information from the $\Sigma(1385)$.
Although the corrections to $R$ from other contamination should be small, 
it is necessary to include them in the calculation.  
There is some probability that contamination for these other channels 
can leak through, and acceptance studies were done for all channels under both 
the $\Lambda\gamma$ and $\Lambda \pi^{0}$ hypotheses.
Results from the acceptance for each hypothesis are shown in Table \ref{acc2}.
The branching ratio must include corrections for the $K^{*+} \to K^+X$ and the
$\Lambda(1405) \to \Sigma\pi$ contamination, as well as a contribution 
to the numerator of $R$ from $\Lambda(1405)\rightarrow \Lambda\gamma$ decay.  
The leakage of the $\Sigma\gamma$ channel is assumed to be small relative 
to the $\Lambda \gamma$ signal.  However, this
channel is still considered in the acceptance studies, see Table \ref{acc2}.

The branching ratio, taking these backgrounds into consideration, is
Eq. (\ref{finalR}).
The $n_\gamma$ ($n_\pi$) terms come directly from the yield of the 
kinematic fits and represent the measured number of photon (pion) candidates.  
A similar notation is used 
so that the pion (photon) channel identifications are denoted 
$A^\Sigma_{\pi}(\Sigma^+\pi^-)$ ($A^\Sigma_{\gamma}(\Sigma^+\pi^-)$), 
where $A^\Sigma_\gamma(\Lambda\pi)$ is the relative leakage of 
the $\Lambda\pi$ channel into the $\Lambda\gamma$ extraction,
$A^\Sigma_\pi(\Lambda\gamma)$ is the relative leakage of the 
$\Lambda\gamma$ channel into the $\Lambda\pi$ extraction, 
$A^{\Sigma}$ is the acceptance strictly for the $\Sigma(1385)$, and
$A^{\Lambda}$ is the acceptance for the $\Lambda(1405)$.

Table \ref{acc2} shows the acceptance terms for the 
other background channels
that are considered in the ratio calculation. 
The table lists three columns sorted by hypothesis: 
$A_{\gamma}$, $A_{\pi}$, and the counts that made all other cuts but 
did not satisfy either the $\gamma$ or $\pi^{0}$ hypothesis, $A_{\gamma\pi}$. 
The latter is used to obtain an estimate of counts for the specific backgrounds listed.

To obtain the values of $N(\Lambda^{*})$ from Eqs. (10)-(13) that must be subtracted
from $n_{\gamma}$ or $n_{\pi}$, the number of observed 
counts of $\Lambda^{*}\to\Sigma^{0}\pi^{0}$ are used with the acceptances from
the third column and the acceptance of the background channel of
interest.  As an example consider $N(\Lambda^{*}\to\Sigma^+\pi^-)$ under the
$\pi$ hypothesis.  The number of observed counts $n(\Sigma^{0}\pi^{0})$ above the $\pi^{0}$ peak
is given by 
\begin{eqnarray}
N(\Lambda^*)=\frac{n_{\Lambda}}{R(\Lambda^*\to\Sigma^{0}\pi^{0})A^{\Lambda}(\Sigma^{0}\pi^{0})}.
\label{eq46}
\end{eqnarray}
The notation $n_{\Lambda}$ here is shorthand for $n(\Sigma^{0}\pi^{0})$, 
while $R(\Lambda^*\to\Sigma^0\pi^0)$ is
the probability that the $\Lambda(1405)$ will decay to $\Sigma^0\pi^0$ 
and $A^{\Lambda}(\Sigma^0\pi^0)$ is the probability that this decay channel 
will be observed after all the applied cuts.  Isospin symmetry is assumed so 
that $R(\Sigma^0\pi^0)=R(\Sigma^+\pi^-)=R(\Sigma^-\pi^+)\approx 1/3$ for 
the $\Lambda(1405)$ decay channels.  An estimate of the number
of counts in the $\pi^{0}$ peak coming from the reaction 
$\Lambda^*\to\Sigma^+\pi^-$, using Eq. (\ref{eq46}), is
\begin{eqnarray}
&N_{\pi}(\Lambda^{*}\to\Sigma^+\pi^-)&\nonumber\\
=&R(\Lambda^{*} \to \Sigma^+\pi^-)A^{\Lambda}_{\pi}(\Sigma^+\pi^-)N(\Lambda^{*})&\nonumber\\
=&A_{\pi}^{\Lambda}(\Sigma^+\pi^-)n_{\Lambda}/A_{\gamma\pi}^{\Lambda}(\Sigma^{0}\pi^{0}).
\end{eqnarray}

A small adjustment is made to ensure that the $\Lambda^*\to \Sigma^+\pi^-$ 
contributions are also included
by adding in the relative acceptance $A^{\Lambda}(\Sigma^+\pi^-)$ to the 
denominator.  These acceptance terms are found by independently using
Monte Carlo for the $\gamma p \to K^{+}\Lambda(1405)\to K^{+}\Sigma^0\pi^0$ 
and $\gamma p \to K^{+}\Lambda(1405)\to K^{+}\Sigma^+\pi^-$ reactions.
The counts that survive all cuts but did not satisfy either 
the $\gamma$ or $\pi^{0}$ hypothesis contribute to $A_{\gamma\pi}$.
The leakage for the $\gamma p \to K^{+}\Lambda(1405)\to K^{+}\Sigma^+\pi^-$ 
channel is very small but is included for
completeness.  The final result is  
\begin{eqnarray}
N_{\pi}(\Lambda^*\to\Sigma^+\pi^-)=\frac{A_{\pi}^{\Lambda}(\Sigma^+\pi^-)n_{\Lambda}}{A_{\gamma\pi}^{\Lambda}(\Sigma^{0}\pi^{0})+A_{\gamma\pi}^{\Lambda}(\Sigma^{+}\pi^{-})}.
\end{eqnarray}
From this example it becomes transparent how to express all other 
associated $\Lambda(1405)$ corrections using only the
observed $n_{\Lambda}$ counts and the corresponding acceptance for this channel.
\end{document}